\documentclass[submission,copyright,creativecommons]{eptcs}
\usepackage{amssymb, amsmath,amsthm}
\usepackage{underscore}
\usepackage{extarrows}
\usepackage{xspace}
\usepackage{tikz}
\usepackage{subcaption}

\providecommand{\event}{ThEdu 2019}

\newtheorem{definition}{Definition}[section]
\newtheorem{lemma}{Lemma}[section]
\theoremstyle{remark}
\newtheorem{example}{Example}[section]

\newcommand{\m}[1]{\mathsf{#1}}
\newcommand{\mc}[1]{\mathcal{#1}}
\newcommand{\Nat}{\mathbb N}
\newcommand{\seq}[2][n]{{#2_1},\dots,{#2_{#1}}}
\newcommand{\OO}{\mathcal O}
\newcommand{\widebar}[1]{\mathop{\overline{#1}}}

\newcommand{\VV}{\mathcal V}
\newcommand{\FF}{\mathcal F}
\newcommand{\TT}{\mathcal T}
\newcommand{\RR}{R}
\newcommand{\EE}{E}
\renewcommand{\AA}{A}
\newcommand{\dc}{\m{dc}}
\renewcommand{\dh}{\m{dh}}
\newcommand{\rc}{\m{rc}}
\newcommand{\AC}{\mathrm{AC}}
\newcommand{\CP}{\m{CP}}

\newcommand{\mr}[1]{\mathrel{#1}}
\newcommand{\scs}{\scriptstyle}
\newcommand{\mirror}[1]{\reflectbox{${#1}$}}
\renewcommand{\L}{\leftarrow}
\newcommand{\R}{\rightarrow}
\newcommand{\LR}{\leftrightarrow}
\newcommand{\J}{\downarrow}
\newcommand{\M}{\uparrow}
\newcommand{\Ra}[1][]{\R^{#1}}
\newcommand{\Rb}[1][]{\R_{#1}}
\newcommand{\Rab}[2][]{\R_{#1}^{#2}}

\newcommand{\Lb}[1][]{\mr{\vphantom{\R}_{#1}{\L}}}
\newcommand{\Lab}[2][]{\mr{\mirror{\R_{\mirror{\scs #1}}^%
{\mirror{\scs #2}}}}}
\newcommand{\LRa}[1][]{\LR^{#1}}
\newcommand{\LRb}[1][]{\LR_{#1}}

\newcommand{\Jab}[2][]{\J_{#1}^{#2}}
\newcommand{\Mab}[2][]{\M_{#1}^{#2}}
\newcommand{\MbR}{\Mab[R]{}}

\newcommand{\RbR}{\Rb[R]}

\newcommand{\RabR}[1]{\Rab[R]{#1}}

\newcommand{\LbR}{\Lb[R]}

\newcommand{\LabR}[1]{\Lab[R]{#1}}
\newcommand{\JbR}{\Jab[R]{}}

\newcommand{\C}[1][]{\LRa[*]_{#1}}
\newcommand{\CbE}{\C[E]}
\newcommand{\CbR}{\C[R]}
\newcommand{\smallparallel}{\raisebox{.07em}{\scalebox{.6}{$\|$}}}
\newcommand{\PR}{\mr{\smash{\xrightarrow%
[\,\smash{\raisebox{1.45ex}{\smallparallel}}\,]{}}}}

\newcommand{\toolname}[1]{\textsf{#1}}
\newcommand{\tool}[1]{\toolname{#1}\protect\xspace}
\newcommand{\TTTT}{\tool{$\textsf{T}\!\protect\raisebox{-1mm}{T}\!\textsf{T}\!%
\protect\raisebox{-1mm}{2}$}}
\newcommand{\TCT}{\tool{$\textsf{T}\!\protect\raisebox{-1mm}{C}\!\textsf{T}\!$}}
\newcommand{\KBCV}{\tool{KBCV}}
\newcommand{\mkbtt}{\tool{mkbTT}}

\newcommand{\lemref}[1]{Lemma~\ref{lem:#1}}

\newcommand{\secref}[1]{Section~\ref{sec:#1}}

\newcommand{\exaref}[1]{Example~\ref{exa:#1}}
\newcommand{\figref}[1]{Figure~\ref{fig:#1}}

\newcommand{\rC}{\m R}
\newcommand{\bC}{\m B}
\newcommand{\gC}{\m G}
\newcommand{\wB}{\circ\vphantom{+}}
\newcommand{\bB}{\bullet\vphantom{+}}
\newcommand{\cons}{\mathrel{:}}
\newcommand{\app}{\mathrel{@}}

\title{Tools in Term Rewriting for Education}
\author{Sarah Winkler
\institute{Universit\`{a} di Verona, Italy}
\email{sarahmaria.winkler@univr.it}
\and
Aart Middeldorp
\institute{University of Innsbruck, Austria}
\email{aart.middeldorp@uibk.ac.at}
}

\makeatletter
\let\thetitle\@title
\makeatother

\def\authorrunning{S. Winkler and A. Middeldorp}
\def\titlerunning{\thetitle}

\begin{document}

\maketitle

\begin{abstract}
Term rewriting is a Turing complete model of computation. When taught to
students of computer science, key properties of computation as well as
techniques to analyze programs on an abstract level are conveyed.
This paper gives a swift introduction to term rewriting and presents
several automatic tools to analyze term rewrite systems which were
developed by the Computational Logic Group at the University of Innsbruck.
These include the termination tool \TTTT, the confluence prover
\tool{CSI}, the completion tools \mkbtt and \KBCV, the complexity tool
\tool{TcT}, the strategy tool \tool{AutoStrat}, as well as \tool{FORT},
an implementation of the decision procedure for the first-order theory for 
a decidable class of rewrite systems. Besides its applications
in research, this software pool has also proved invaluable for teaching,
e.g., in multiple editions of the International Summer School on
Rewriting.
\end{abstract}

\section{Introduction}

Rewriting is a pervasive concept in mathematics, computer science, and
other areas: Simplification of expressions constitutes rewriting, the
execution of
a program can be seen as a rewrite sequence on program states, and in fact
probably almost any development according to a set of fixed rules can be
considered rewriting.
In \emph{term rewriting}, we assume that the objects which are rewritten
are terms. This yields a powerful formalism which is crucial for
simplification in automated theorem proving, it provides tools to analyze
security protocols, it can be used to model the development of RNA
structures, but it is also a versatile method in program verification, to
name only a few application areas.
In fact, term rewriting is a Turing-complete model of computation, and
provides methods
to investigate important properties of computation and
simplification processes on an abstract level~\cite{BN98,TeReSe}. 

\smallskip

This includes ubiquituous properties related to termination, determinism,
and complexity. As a simple but powerful model of computation, term
rewriting can in particular also convey program analysis
on an abstract level to students of computer science and related fields.
We illustrate some properties by means of a simple example.

\begin{example}[Coffee Bean Game~\cite{DP01}]
\label{exa:beans}
Coffee beans come in two kinds called black ($\bB$) and white ($\wB$). A
two-player game starts with a random sequence of black and white beans. In
a move, a player must take two adjacent beans and put back one bean,
according to the following set of rules $\RR_1$:
\begin{xalignat*}{4}
\bB~\bB &\to \wB &
\wB~\wB &\to \wB &
\bB~\wB &\to \bB &
\wB~\bB &\to \bB
\end{xalignat*}
The player who puts the last white bean wins.
For instance, the following is a valid game:
\begin{center}
$\begin{array}{c}
\bB~\underline{\wB~\wB}~\bB~\wB~\bB~\bB~\wB~\wB~\bB~\wB~\wB~\bB~\bB~\wB
\\[-.5ex]
\bB~\wB~\bB~\wB~\bB~\bB~\wB~\wB~\bB~\wB~\underline{\wB~\bB}~\bB~\wB
\\[-.5ex]
\bB~\wB~\bB~\wB~\bB~\bB~\underline{\wB~\wB}~\bB~\wB~\wB~\bB~\wB
\\[-.5ex]
\bB~\wB~\bB~\wB~\bB~\bB~\wB~\bB~\wB~\underline{\wB~\bB}~\wB
\\[-.5ex]
\bB~\underline{\wB~\bB}~\wB~\bB~\bB~\wB~\bB~\wB~\bB~\wB
\\[-.5ex]
\bB~\bB~\wB~\bB~\bB~\wB~\bB~\wB~\underline{\bB~\wB}
\\[-.5ex]
\bB~\bB~\wB~\bB~\bB~\wB~\bB~\underline{\wB~\bB}
\\[-.5ex]
\bB~\bB~\wB~\bB~\bB~\wB~\underline{\bB~\bB}
\\[-.5ex]
\bB~\bB~\wB~\bB~\bB~\underline{\wB~\wB}
\\[-.5ex]
\bB~\bB~\wB~\bB~\underline{\bB~\wB}
\\[-.5ex]
\bB~\bB~\wB~\underline{\bB~\bB}
\\[-.5ex]
\bB~\bB~\underline{\wB~\wB}
\\[-.5ex]
\bB~\underline{\bB~\wB}
\\[-.5ex]
\underline{\bB~\bB}
\\[-.5ex]
\wB
\end{array}$
\end{center}
In this case the player who started lost, since the last white bean was
put in the 14th move. A number of interesting questions can be asked
about such a game:
Which moves should the respective players perform to win?
Are there game states which are equivalent in the sense that they offer
the same opportunities to each of the players?
In short, is there a winning strategy for one of the players?
While it is obvious that the above game terminates, is this still the case
for the modified game using the rules $\RR_2$:
\begin{xalignat*}{4}
\bB~\bB &\to \wB~\wB~\wB~\wB &
\wB~\wB &\to \wB &
\bB~\wB &\to \wB~\wB~\wB~\bB &
\wB~\bB &\to \bB
\end{xalignat*}
and if yes, how many steps are needed?
\end{example}

This paper advocates rewriting to answer these questions and many others
that we will motivate by examples.
As manual analysis of term rewrite systems often turns out to be tedious,
a variety of tools has been developed in the last two decades which
perform powerful analysis tasks automatically.
We here focus on tools that have been developed at the Computational
Logic group at the University of Innsbruck since these are the tools we
are 
most familiar with.

\smallskip

This paper gives a concise introduction to term rewriting. We introduce
some of the most widely investigated properties of term rewrite systems,
and motivate their relevance by examples from different domains.
Rather than elaborating the often complicated methods developed to analyze
these properties, we show how tools can effectively be used to inspect
term rewrite systems automatically. In this spirit we discuss
termination (\secref{termination}), confluence (\secref{confluence}),
completion as a means to decide the validity problem (\secref{completion}),
the first order theory of rewriting (\secref{first-order theory}),
evaluation strategies (\secref{strategies}), and
derivational complexity (\secref{complexity}).
We conclude in \secref{conclusion} with remarks on current research.

\section{Preliminaries}

We assume basic familiarity with term rewriting~\cite{BN98,TeReSe},
but recall some key notions and notation.
Given a signature $\FF$ and a set of variables $\VV$, we 
consider the set of terms $\TT$ built up from $\FF$ and $\VV$.
\emph{Positions} are strings of positive integers which are
used to address subterms. We write $t|_p$ for the subterm of $t$
at position $p$ and $t[u]_p$ denotes the term that is obtained from
$t$ by replacing its subterm $t|_p$ with $u$.
A \emph{substitution} $\sigma$ is a mapping
from variables to terms such that $\sigma(x) \neq x$ for only 
finitely many $x$.
An \emph{equation} is a pair of terms $s \approx t$, and a
\emph{rewrite rule} is a pair of terms denoted as $\ell \to r$
such that $\ell \notin \VV$ and all variables in $r$ also occur in
$\ell$. An equational system (ES) is a set of equations,
while a \emph{term rewrite system} (TRS) refers to a set of rewrite
rules.

\smallskip

The \emph{rewrite relation} induced by a TRS $\RR$ is defined as
$s \RbR t$ if and only if
$s|_p = \ell\sigma$ and 
$t = s[r\sigma]_p$ for some position $p$, substitution $\sigma$, and 
rewrite rule $\ell \to r$ in $\RR$. The relations
$\LRb[R]$, $\RabR{+}$, and $\RabR{*}$ denote the
symmetric, transitive, and reflexive transitive closure of 
$\RbR$, respectively, while the reflexive, symmetric, and transitive
closure of $\RbR$ is denoted $\CbR$ and called \emph{conversion}.
Two terms $s$ and $t$ are \emph{convertible} if there exists a
conversion $s \CbR t$.
We further use $\JbR$ as abbreviation for the
\emph{joinability relation} $\LabR{*} \cdot \RabR{*}$
and $\uparrow_\RR$ as abbreviation for the
\emph{meetability relation} $\RabR{*} \cdot \LabR{*}$.
Here $\cdot$ denotes relation composition.
A \emph{normal form} with respect to a TRS $\RR$ is a term $t$ such
that there is no term $s$ with $t \RbR s$. We also write 
$u \RabR{!} t$ if $u \RabR{*} t$ and $t$ is a normal form.

\smallskip

Some further concepts will be introduced in later sections
when they are needed.

\section{Termination}
\label{sec:termination}

Termination is very often a desired feature of rewrite systems,
and thus one of the most studied properties.

\begin{definition}
A \textup{TRS} $\RR$ is terminating if there is no infinite rewrite
sequence $t_0 \RbR t_1 \RbR t_2 \RbR \cdots$.
\end{definition}

\begin{example}
\label{exa:beans2}
We revisit \exaref{beans} from the introduction.
It is obvious that the TRS $\RR_1$ terminates since the number of beans
decreases by one with every move. Though the case of $\RR_2$ is less
obvious, it turns out that also this TRS terminates. Many different
techniques can be harnessed to show this. Here we use this example to
illustrate a popular technique to show termination based on
\emph{interpretations}.

\smallskip

Suppose we take as carrier set the natural numbers and use
$\wB_\AA(x) = x + 1$ and $\bB_\AA(x) = 4x + 1$ as interpretations. The
terms in the four rewrite rules 
\begin{xalignat*}{4}
\bB(\bB(x)) &\to \wB(\wB(\wB(\wB(x)))) &
\wB(\wB(x)) &\to \wB(x) &
\bB(\wB(x)) &\to \wB(\wB(\wB(\bB(x))))  &
\wB(\bB(x)) &\to \bB(x)
\end{xalignat*}
then correspond to the following polynomials, where independent of the
value of $x$ the left-hand side is always greater than the right-hand side:
\begin{xalignat*}{4}
16x + 5 &>_\Nat x + 4 &
x + 2 &>_\Nat x + 1 &
4x + 5 &>_\Nat x + 4 &
4x+2 &>_\Nat 4x+1
\end{xalignat*}
Since every rewrite step results in a strict decrease, and
$>_\Nat$ is a well-founded order, this linear polynomial interpretation
shows that no infinite rewrite sequence can possibly exist.
\end{example}

\smallskip

\TTTT~\cite{KSZM09} is a tool to show termination of TRSs, available both
via a web interface and as a stand\-alone executable.%
\footnote{\url{http://cl-informatik.uibk.ac.at/software/ttt2/}}
It is beyond the scope of this
paper to describe all implemented techniques; we only mention that a great
variety of approaches is supported, including different term orders,
interpretations over various domains, modularization of termination
problems according to the powerful dependency pair framework, and
numerous specialized routines. The tool also provides support for relative
termination, as well as means to show termination with respect to
strategies (see \secref{strategies}) and nontermination.
Upon success, \TTTT outputs all details of the (non)termination 
proof such as interpretations, parameters of orderings, or a 
counterexample in case termination was disproved.
This helps students (as well as researchers) to understand the 
result.

\smallskip

\begin{figure}[tb]
\centering
\includegraphics[width=.98\linewidth]{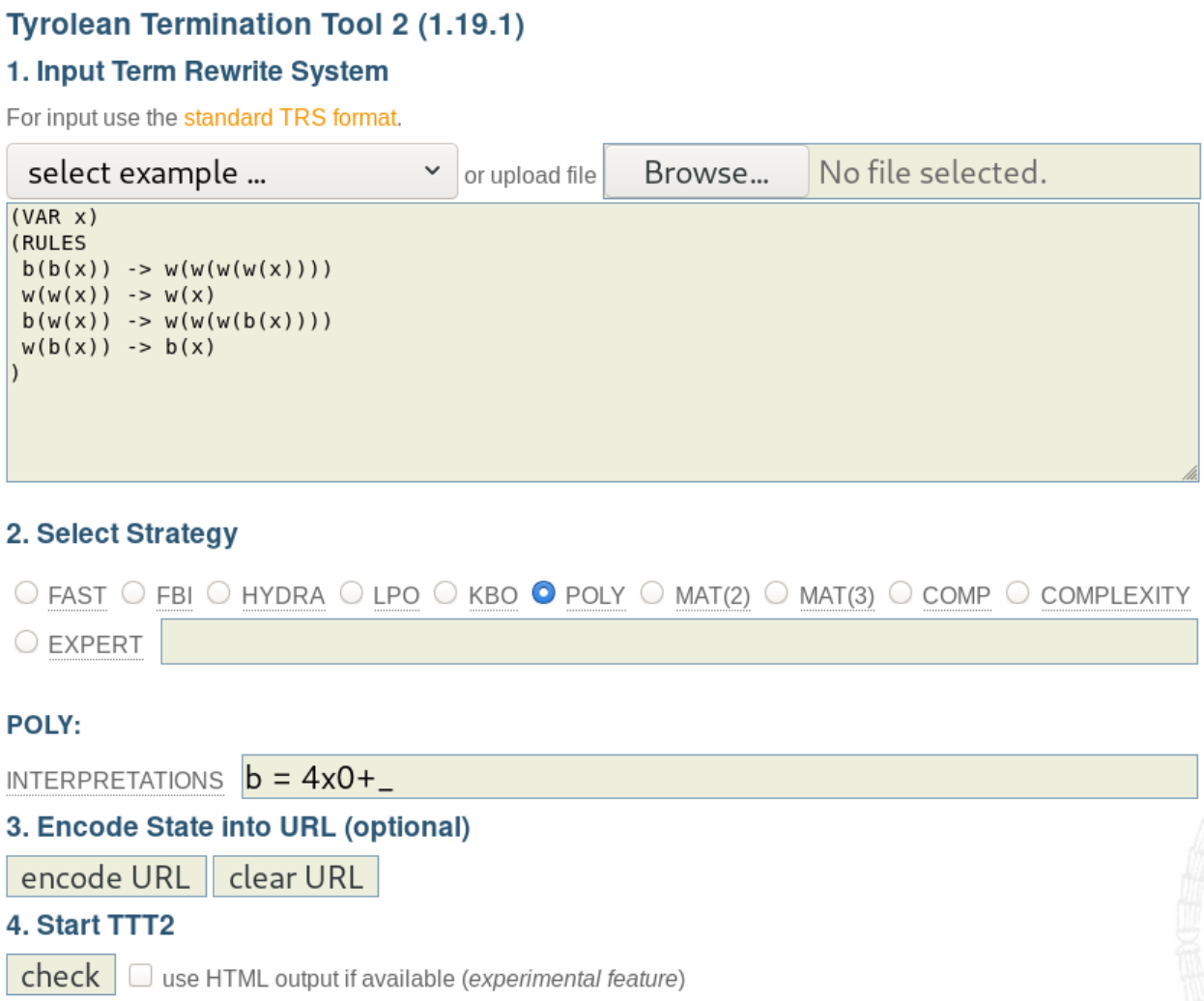}
\caption{The web interface of \TTTT.}
\label{fig:ttt2 screenshot}
\end{figure}

Specialized support for teaching was added recently \cite{SS18}.
We mention the encoding of the state of the web interface into a URL,
which allows examples that are used for teaching to be directly loaded
into the web interface by a simple mouse click from the slides. This
avoids time-consuming and error-prone manipulations during a lecture or
talk. To illustrate this,
\href{http://colo6-c703.uibk.ac.at/ttt2/web/?problem=(VAR\%20x)\%0A(RULES%
\%0A\%20b(b(x))\%20-\%3E\%20w(w(w(w(x))))\%0A\%20w(w(x))\%20-\%3E\%20w(x)%
\%0A\%20b(w(x))\%20-\%3E\%20w(w(w(b(x))))\%0A\%20w(b(x))\%20-\%3E\%20b(x)%
\%0A)&strategy=poly&template=b\%20\%3D\%204x0\%2B_}{clicking here}
opens the web interface of \TTTT in a browser with the above bean rules
and a partial polynomial interpretation, indicating that we look for
an interpretation with $\bB_\AA(x) = 4x + c$ for some constant $c$. A
screenshot is shown in \figref{ttt2 screenshot}. Guiding
termination methods by providing some of the parameters is also supported
for the Knuth--Bendix order (KBO), the lexicographic path order 
(LPO), and  matrix interpretations, since these are the termination
methods taught in the bachelor course on term rewiting at the University
of Innsbruck.
This feature is useful for students in multiple respects:
Sometimes exercises demand to complete a given partial interpretation,
in other cases an interpretation of a particular shape is demanded;
in both cases students can check their solutions with this functionality.
But it also helps them to refine their own incomplete solutions, and 
can be used to show that, for instance, a certain precedence relation
between two function symbols does \emph{not} work for LPO or KBO.

\smallskip

We conclude this section with another example where termination is
less obvious.

\begin{example}[Battle of Hydra and Hercules]
The mythological monster Hydra is a dragon-like creature with multiple
heads. Whenever Hercules in his fight chops off a head, more and more new
heads can grow instead, since the beast gets increasingly angry. 
Here we model a Hydra as an unordered tree. If Hercules cuts off a
leaf corresponding to a head, the tree is modified in the following way:
If the cut-off node $h$ has a grandparent $n$, then the branch from $n$ to
the parent of $h$ gets multiplied, where the number of copies depends on
the number of decapitations so far. Hydra dies if there are no heads left,
in that case Hercules wins. The following sequence shows an example
fight:
\begin{center}
\begin{tikzpicture}[level distance=5mm,
  level 1/.style={sibling distance=6mm},
  level 2/.style={sibling distance=4mm},]
\tikzstyle{inner}=[minimum size=1.8mm,inner sep=0pt,draw,fill=white,
  shape=circle]
\tikzstyle{leaf}=[minimum size=1.8mm,inner sep=0pt,fill=black,
  shape=circle]
\tikzstyle{cut}=[minimum size=4mm,inner sep=0pt,fill=none,shape=circle,
  draw=blue]
\tikzstyle{stage}=[scale=1]
\node[inner] at (0,0) {} child {
  node[leaf] {}
  } child {
  node[inner] {} child {
    node[inner] {} child {
      node[inner] {} child {
        node[leaf] (A) {}
      }
      } child {
      node[inner] {} child {
        node[leaf] {} 
        }
      }
    }
  } child {
  node[inner] {} child {
    node[leaf] {}
  }
  };
\node[cut] at (A) {};
\node[stage] at (0,-2.7) {1};
\node[inner] at (3.2,0) {} child {
  node[leaf] {}
  } child {
  node[inner] {} child {
    node[inner] {} child {
      node[leaf] {}
      } child {
      node[leaf] (B) {}
      } child {
      node[inner] {} child {
        node[leaf] {}
        }
      }
    }
  } child {
  node[inner] {} child {
    node[leaf] {}
  }
  };
\node[cut] at (B) {};
\node[stage] at (3.2,-2.7) {2};
\begin{scope}[level distance = 5mm,
  level 1/.style={sibling distance=11mm},
  level 2/.style={sibling distance=6mm},
  level 3/.style={sibling distance=3mm},]
\node[inner] at (6.4,0) {} child {
  node[leaf] {}
  } child {
  node[inner] {} child {
    node[inner] {} child {
      node[leaf] {}
      } child {
      node[inner] {} child {
        node[leaf] {}
        }
      }
    }child {
    node[inner] {} child {
      node[leaf] {}
      } child {
      node[inner] {} child {
        node[leaf] {}
        }
      }
    }child {
    node[inner] {} child {
      node[leaf] {}
      } child {
      node[inner] {} child {
        node[leaf] {}
        }
      }
    }
  } child {
  node[inner] {} child {
    node[leaf] (C) {}
  }
  };
\node[cut] at (C) {};
\node[stage] at (6.4,-2.7) {3};
\end{scope}
\begin{scope}[level distance = 5mm,
  level 1/.style={sibling distance=5mm},
  level 2/.style={sibling distance=6.5mm},
  level 3/.style={sibling distance=2.8mm},]
\node[inner] at (9.6,0) {} child {
  node[leaf] {}
  } child {
  node[leaf] {}
  } child {
  node[inner] {} child {
    node[inner] {} child {
      node[leaf] {}
      } child {
      node[inner] {} child {
        node[leaf] {}
        }
      }
    }child {
    node[inner] {} child {
      node[leaf] {}
      } child {
      node[inner] {} child {
        node[leaf] {}
        }
      }
    }child {
    node[inner] {} child {
      node[leaf] {}
      } child {
      node[inner] {} child {
        node[leaf] {}
        }
      }
    }
  } child {
  node[leaf](D)  {}
  } child {
  node[leaf] {}
  } child {
  node[leaf] {}
  };
\node[cut] at (D) {};
\node[stage] at (9.6,-2.7) {4};
\end{scope}
\begin{scope}[level distance = 5mm,
  level 1/.style={sibling distance=5mm},
  level 2/.style={sibling distance=6.5mm},
  level 3/.style={sibling distance=2.8mm},]
\node[inner] at (12.8,0) {} child {
  node[leaf] {}
  } child {
  node[leaf] {}
  } child {
  node[inner] {} child {
    node[inner] {} child {
      node[leaf] {}
      } child {
      node[inner] {} child {
        node[leaf] {}
        }
      }
    }child {
    node[inner] {} child {
      node[leaf] {}
      } child {
      node[inner] {} child {
        node[leaf] {}
        }
      }
    }child {
    node[inner] {} child {
      node[leaf] {}
      } child {
      node[inner] {} child {
        node[leaf] {}
        }
      }
    }
  } child {
  node[leaf] {}
  } child {
  node[leaf] {}
  };
\node[stage] at (12.8,-2.7) {5};
\end{scope}
\end{tikzpicture}
\end{center}
Though the number of heads can grow considerably in one step, it turns out
that the fight always terminates, and Hercules will win independent of
his strategy. This can be shown by an argument based on ordinals.
Touzet modeled this process as a TRS $\mc H$~\cite{To98}.
However, derivations may get very long: their length cannot be described
by a multiple recursive function in the size of the initial monster.
This was one of the reasons that Touzet's TRS remained out of reach for
automatic tools for more than a decade, until
\emph{ordinal interpretations} were developed to deal with such
systems~\cite{ZWM15}.
Nowadays, \TTTT can show termination of $\mc H$ automatically, using an
implementation of this technique.
\end{example}

\section{Confluence}
\label{sec:confluence}

In many applications, it is of interest to know whether the process
described by a TRS satisfies properties related to determinism. For
instance, in \exaref{beans} one would like to know whether different
strategies of the players lead to different results. The most studied
property in rewriting in this context is confluence, defined below.

\begin{definition}
A \textup{TRS} $\RR$ is \emph{locally confluent} if
${\LbR \cdot \RbR} \subseteq {\JbR}$,
and \emph{confluent} if ${\MbR} \subseteq {\JbR}$ holds.
\end{definition}

According to a famous result by Newman~\cite{N42}, these two
properties coincide if a TRS is terminating.

\begin{lemma}
\label{lem:newman}
A terminating and locally confluent \textup{TRS} is confluent.
\qed
\end{lemma}

The definiton of confluence imposes a condition on all \emph{peaks}, i.e.,
rewrite sequences of the form $\LabR{*} \cdot \RabR{*}$, of which
there might be infinitely many, in addition to the fact that
$\RabR{*}$ is in general undecidable. 
Fortunately, it is known that a TRS $\RR$ is locally confluent if
all its critical pairs $\CP(\RR)$ are joinable, of which there are only
finitely many.
Thus it turns out that it suffices to consider finitely many
peaks for local confluence, as expressed by \lemref{cp} below.

\begin{definition}
Let $\ell_1 \to r_1$ and $\ell_2 \to r_2$ be renamings of rewrite rules
in $\RR$ without common variables, such that the following conditions are
satisfied:
\begin{itemize}
\item
$p$ is a non-variable position in $\ell_2$,
\item
$\sigma$ is a most general unifier of $\ell_2|_p$  and $\ell_1$, and
\item
if $p = \epsilon$ then $\ell_1 \to r_1$ and $\ell_2 \to r_2$ are not
variants.
\end{itemize}
The triple $\langle \ell_1 \to r_1, \,p,\, \ell_2 \to r_2\rangle$
constitutes a \emph{critical overlap}, and
$\ell_2\sigma[r_1\sigma]_p \approx r_2\sigma$ is a
\emph{critical pair} of $\RR$.
\end{definition}

Two rewrite steps $s \LbR \cdot \RbR t$ are said to form a
\emph{critical peak} if $s \approx t$ is a critical pair, and the
critical pair is called joinable if $s \JbR t$.
In the sequel we denote the set of all critical pairs of $\RR$ by
$\CP(\RR)$. The following lemma explains the importance of
critical pairs for local confluence.

\begin{lemma}[Critical Pair Lemma~\cite{KB70}]
\label{lem:cp}
Consider a \textup{TRS} $\RR$ and terms $s$ and $t$. If there is a peak
$s \LbR \cdot \RbR t$ then $s \JbR t$ or $s \LRb[\CP(\RR)] t$.
\qed
\end{lemma}

In connection with Newman's Lemma, the Critical Pair Lemma implies that
confluence is decidable for terminating systems. This result can be used
to investigate determinism of the bean game given in the introduction.

\begin{example}
\label{exa:beans confluence}
We investigate confluence of the TRS $\RR_1$ from \exaref{beans}. Since
the TRS is terminating, confluence and local confluence coincide. We thus
analyze the critical pairs of $\RR_1$.
The following eight diagrams show all critical peaks of $\RR$:
\newcommand{\osp}{\!\!\!\widebar{\phantom{\wB}}\!\!\!}
\newcommand{\TRIVIAL}[3]{
\begin{tikzpicture}[baseline=(u.base),scale=.6,xscale=.9,
every node/.style={scale=.7}]
\begin{scope}
\node (u) {$\widebar{#1}\osp\underline{\widebar#2}$};
\node (s) at (-1,-1) {$#3$};
\node (t) at (1,-1) {$#3$};
\draw[->] (u) to (s);
\draw[->] (u) to (t);
\end{scope}
\end{tikzpicture}
}
\newcommand{\TRIVIALs}[3]{
\begin{tikzpicture}[baseline=(u.base),scale=.7,xscale=.9,
every node/.style={scale=.7}]
\begin{scope}
\node (u) {$\widebar{#1}\osp\underline{\widebar#2}$};
\node (s) at (-1,-1) {$#3$};
\node (t) at (1,-1) {$#3$};
\draw[->] (u) to (s);
\draw[->] (u) to (t);
\end{scope}
\end{tikzpicture}
}
\newcommand{\DIAMOND}[5]{
\begin{tikzpicture}[baseline=(u.base),scale=.6,xscale=.9,
every node/.style={scale=.7}]
\begin{scope}
\node (u) {$\widebar{#1}\osp\underline{\widebar#2}$};
\node (s) at (-1,-1) {$#3$};
\node (t) at (1,-1) {$#4$};
\draw[->] (u) to (s);
\draw[->] (u) to (t);
\node (v) at (0,-2) {$#5$};
\draw[->] (s) to (v);
\draw[->] (t) to (v);
\end{scope}
\end{tikzpicture}
}
\newcommand{\DIAMONDs}[5]{
\begin{tikzpicture}[baseline=(u.base),scale=.7,xscale=.9,
every node/.style={scale=.7}]
\begin{scope}
\node (u) {$\widebar{#1}\osp\underline{\widebar#2}$};
\node (s) at (-1,-1) {$#3$};
\node (t) at (1,-1) {$#4$};
\draw[->] (u) to (s);
\draw[->] (u) to (t);
\node (v) at (0,-2) {$#5$};
\draw[->] (s) -- (v) node[midway,sloped,below] {$\scriptstyle *$};
\draw[->] (t) -- (v) node[midway,sloped,below] {$\scriptstyle *$};
\end{scope}
\end{tikzpicture}
}
\begin{align*}
&\TRIVIAL{\wB}{\wB\wB}{\wB\wB} &
&\TRIVIAL{\wB}{\wB\bB}{\wB\bB} &
&\DIAMOND{\wB}{\bB\wB}{\wB\bB}{\bB\wB}{\bB} &
&\DIAMOND{\wB}{\bB\bB}{\wB\wB}{\bB\bB}{\wB} &
&\TRIVIAL{\bB}{\wB\wB}{\bB\wB} &
&\TRIVIAL{\bB}{\wB\bB}{\bB\bB} &
&\DIAMOND{\bB}{\bB\wB}{\bB\bB}{\wB\wB}{\wB} &
&\DIAMOND{\bB}{\bB\bB}{\bB\wB}{\wB\bB}{\bB}
\end{align*}
In each of these local peaks, the rewrite steps either lead to the same
result, or the resulting two terms have a common reduct that is reached
in a single step from both.
Thus $\RR_1$ is locally confluent by \lemref{cp}, and confluent by
\lemref{newman} since $\RR_1$ is terminating.
This implies that the result of the game only depends on the initial
configuration. A similar analysis applies to the
TRS $\RR_2$ of \exaref{beans}, although finding a common reduct
requires more steps:
\begin{align*}
&\TRIVIALs{\wB}{\wB\wB}{\wB\wB} &
&\DIAMONDs{\bB}{\bB\bB}{\wB\wB\wB\wB\bB}{\bB\wB\wB\wB\wB}{\bB} &
&\DIAMONDs{\bB}{\bB\wB}{\bB\wB\wB\wB\bB}{\wB\wB\wB\wB\wB}{\wB} &
&\DIAMONDs{\bB}{\wB\wB}{\bB\wB}{\wB\wB\wB\bB\wB}{\bB} \\
&\TRIVIALs{\wB}{\wB\bB}{\wB\bB} &
&\DIAMONDs{\wB}{\bB\bB}{\wB\wB\wB\wB\wB}{\bB\bB}{\wB} &
&\DIAMONDs{\bB}{\wB\bB}{\bB\bB}{\wB\wB\wB\bB\bB}{\wB} &
&\DIAMONDs{\wB}{\bB\wB}{\wB\wB\wB\wB\bB}{\bB\wB}{\bB}
\end{align*}
\end{example}

For systems that are non-terminating, joinability of critical pairs
is insufficient for confluence. By forbidding criticial pairs and
imposing the condition that left-hand sides of rules do not contain
repeated variables (left-linearity), confluence is guaranteed~\cite{R73}.
This syntactic criterion is called \emph{orthogonality}. We give an
example.

\begin{example}
\label{exa:eratosthenes}
The following TRS $\RR$ models a functional program to enumerate prime
numbers:
\begin{align*}
\m{primes} &\to \m{sieve}(\m{from}(\m{s}(\m{s}(\m{0})))) &
\m{sieve}(\m{0}:y) &\to \m{sieve}(y) \\
\m{from}(n) &\to n:\m{from}(\m{s}(n)) &
\m{sieve}(\m{s}(n):y) &\to \m{s}(n):\m{sieve}(\m{filter}(n,y,n)) \\
\m{take}(\m{0},y) &\to \m{nil} &
\m{filter}(\m{0},x:y,m) &\to \m{0}:\m{filter}(m,y,m) \\
\m{take}(\m{s}(n),x:y) &\to x:\m{take}(n,y) &
\m{filter}(\m{s}(n),x:y,m) &\to x:\m{filter}(n,y,m)
\end{align*}
It does not terminate as it can, for instance, exhibit the
sequence $\m{from}(\m{0}) \to \m{0}:\m{from}(\m{s}(\m{0}))
\to \m{0}:\m{s}(\m{0}):\m{from}(\m{s}(\m{s}(\m{0}))) \to \dots$.
However, if the corresponding code is executed using lazy evaluation
then non-termination is not a problem for the program.
Since the TRS is orthogonal, confluence does hold. As a consequence,
every term has at most one normal form. For instance,
a call $\m{take}(\m{s}(\m{s}(\m{0})),\m{primes})$ evaluates
to the (unique) normal form
$\m{s}(\m{s}(\m{0})):\m{s}(\m{s}(\m{s}(\m{0}))):\m{nil}$,
representing the list consisting of the first two prime numbers.
\end{example}

In a first course on term rewriting the two sufficient
conditions described above are typically taught to students and every
confluence tool supports these techniques. \tool{CSI}~\cite{NFM17,ZFM11}
is developed in Innsbruck. It is built on top of \TTTT and available
via a web interface and as a stand-alone executable.%
\footnote{\url{http://cl-informatik.uibk.ac.at/software/csi/}}
The web interface is less elaborate than the one of \TTTT. One reason
for this is that the basic sufficient conditions do not have
parameters that need to be instantiated.
But just like \TTTT also \tool{CSI} outputs all details of the
(non)confluence proof to make proof reconstruction for users as easy
as possible.

\smallskip

\begin{figure}[tb]
\centering
\includegraphics[width=.98\linewidth]{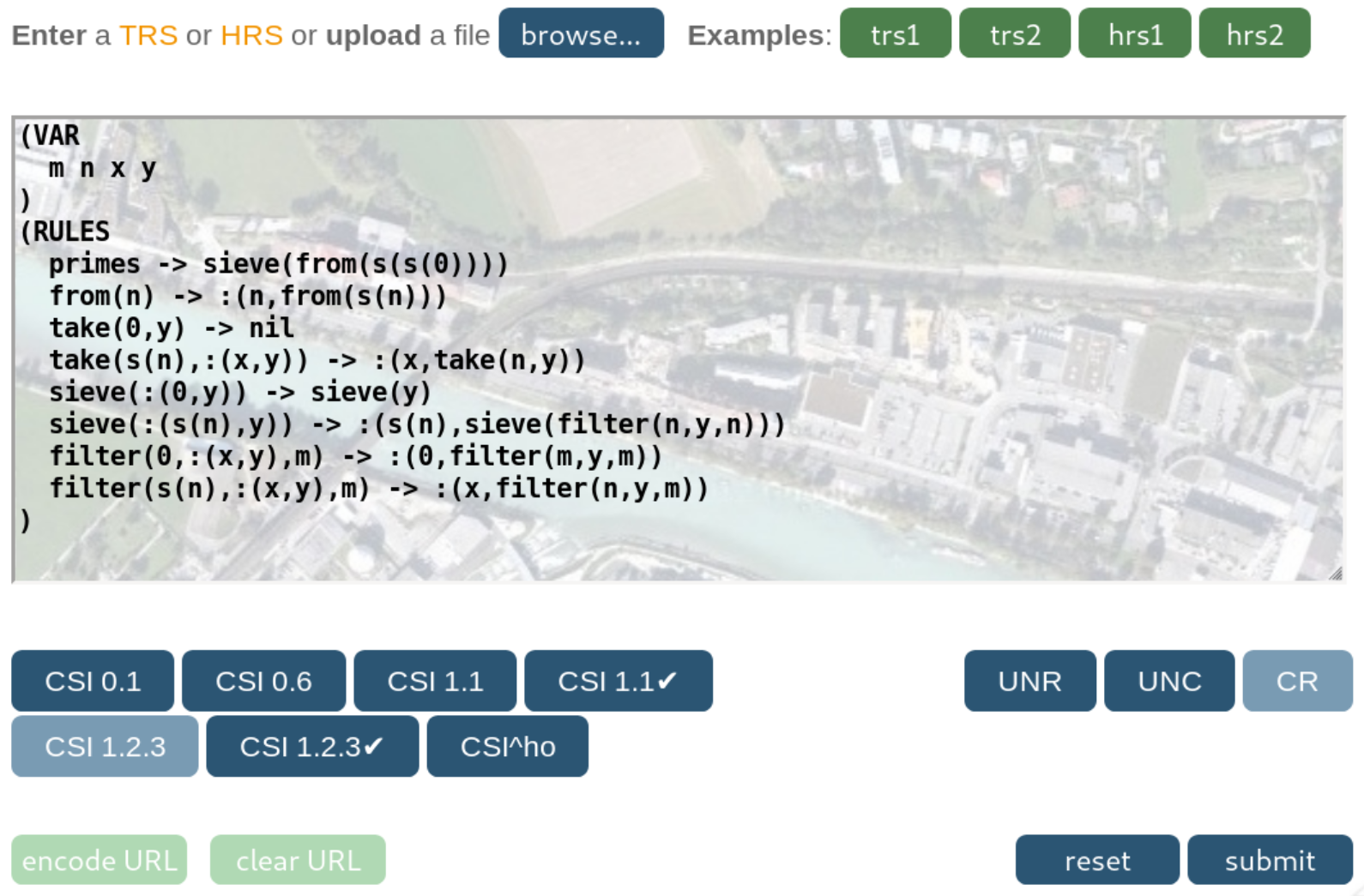}
\caption{The web interface of \tool{CSI}.}
\label{fig:csi screenshot}
\end{figure}

For this paper we added URL encoding. As a consequence, a
\href{http://colo6-c703.uibk.ac.at/csi/index.php?%
problem=(VAR\%0A\%20\%20m\%20n\%20x\%20y\%0A)\%0A(RULES\%0A\%20\%20primes%
\%20-\%3E\%20sieve(from(s(s(0))))\%0A\%20\%20from(n)\%20-\%3E\%20\%3A(n%
\%2Cfrom(s(n)))\%0A\%20\%20take(0\%2Cy)\%20-\%3E\%20nil\%0A\%20\%20take%
(s(n)\%2C\%3A(x\%2Cy))\%20-\%3E\%20\%3A(x\%2Ctake(n\%2Cy))\%0A\%20\%20%
sieve(\%3A(0\%2Cy))\%20-\%3E\%20sieve(y)\%0A\%20\%20sieve(\%3A(s(n)\%2Cy))%
\%20-\%3E\%20\%3A(s(n)\%2Csieve(filter(n\%2Cy\%2Cn)))\%0A\%20\%20filter%
(0\%2C\%3A(x\%2Cy)\%2Cm)\%20-\%3E\%20\%3A(0\%2Cfilter(m\%2Cy\%2Cm))\%0A%
\%20\%20filter(s(n)\%2C\%3A(x\%2Cy)\%2Cm)\%20-\%3E\%20\%3A(x\%2Cfilter(n%
\%2Cy\%2Cm))\%0A)&property=CR&version=csi123}%
{mouse click} suffices to preload the TRS of \exaref{eratosthenes}.
The result is shown in \figref{csi screenshot}.

\smallskip

Numerous other techniques
have been developed for ensuring confluence and related properties
like unique normal forms, some of which are occasionally taught in
advanced courses on rewriting. \tool{CSI} is not the only confluence
tool around. All tools for confluence and related properties that
participate in the yearly Confluence Competition (CoCo)~\cite{MNS19} are
available via CoCoWeb,%
\footnote{\url{http://cocoweb.uibk.ac.at/}}
a convenient web interface that provides a single entry point to all
tools~\cite{HNM18}.

\section{Completion}
\label{sec:completion}

Before mentioning any relevant theory, we provide some examples.
The first one appeared as a contest in a Dutch popular science
magazin~\cite{NWT05}.

\begin{example}
\label{exa:genes}
Genetic engineers in a (hypothetical) research lab want to create cows
that produce cola instead of milk. To that end they plan to transform the
DNA of the milk gene represented by the sequence $\m{TAGCTAGCTAGCT}$ in
every fertilized egg into the cola gene, i.e., the sequence
$\m{CTGACTGACT}$. The research group already developed techniques to
perform the following DNA transformations:
\begin{align*}
\m{TCAT} &\leftrightarrow \m{T} &
\m{GAG}  &\leftrightarrow \m{AG} &
\m{CTC}  &\leftrightarrow \m{TC} &
\m{AGTA} &\leftrightarrow \m{A} &
\m{TAT}  &\leftrightarrow \m{CT}
\end{align*}
However, recently it has been discovered that the mad cow disease is
caused by a retrovirus with the DNA sequence $\m{CTGCTACTGACT}$.
Could it happen that accidentally cows with this virus are created?
\end{example}

\begin{example}[Chameleon Island~\cite{DP01}]
\label{exa:chameleons}
A colony of chameleons on a remote island consists of 20 red, 18 blue,
and 16 green individuals which continuously walk around.
Whenever two chameleons of different color meet, both change to the third
color, i.e., they change according to the following rewrite rules:
\begin{xalignat*}{3}
\rC \cdot \gC &\to \bC \cdot \bC &
\bC \cdot \rC &\to \gC \cdot \gC &
\gC \cdot \bC &\to \rC \cdot \rC \\
\gC \cdot \rC &\to \bC \cdot \bC &
\rC \cdot \bC &\to \gC \cdot \gC &
\bC \cdot \gC &\to \rC \cdot \rC
\end{xalignat*}
Some time passes during which no chameleons are born or die nor do any
enter or leave the colony. Is it possible that after this period, all 54
chameleons are of the same color?
\end{example}

Both of these examples can be seen as instances of the
\emph{validity problem}: Given a set of rewrite rules $\RR$ and two
terms $s$ and $t$, does $s \CbR t$ hold? While this problem is
undecidable in general, \emph{Knuth--Bendix completion}~\cite{KB70}
is a method to solve some instances.

\begin{definition}
A \textup{TRS} is \emph{complete} if it is confluent and terminating. A
\emph{completion procedure} takes as input an \textup{ES} $\EE$ and
attempts to generate a complete \textup{TRS} $\RR$ such that
${\CbE} = {\CbR}$.
\end{definition}

If successful, the resulting TRS $\RR$ can be used to decide the validity
problem: by the properties of a completion procedure and because $\RR$ is
complete the following equivalences hold:
\[
{\CbE} ~=~ {\CbR} ~=~ {\RabR{!} \cdot \LabR{!}}
\]
Therefore, for any two terms $s$ and $t$, $s \CbE t$
if and only if $s$ and $t$ have the same $\RR$-normal form, which is
unique since $\RR$ is confluent.
However, since the validity problem is undecidable, completion does not 
always succeed: it may also fail if some equations cannot be
appropriately processed, or run indefinitely.

\smallskip

\begin{figure}[tb]
\centering
\begin{subfigure}[c]{.3\linewidth}
\includegraphics[width=5cm]{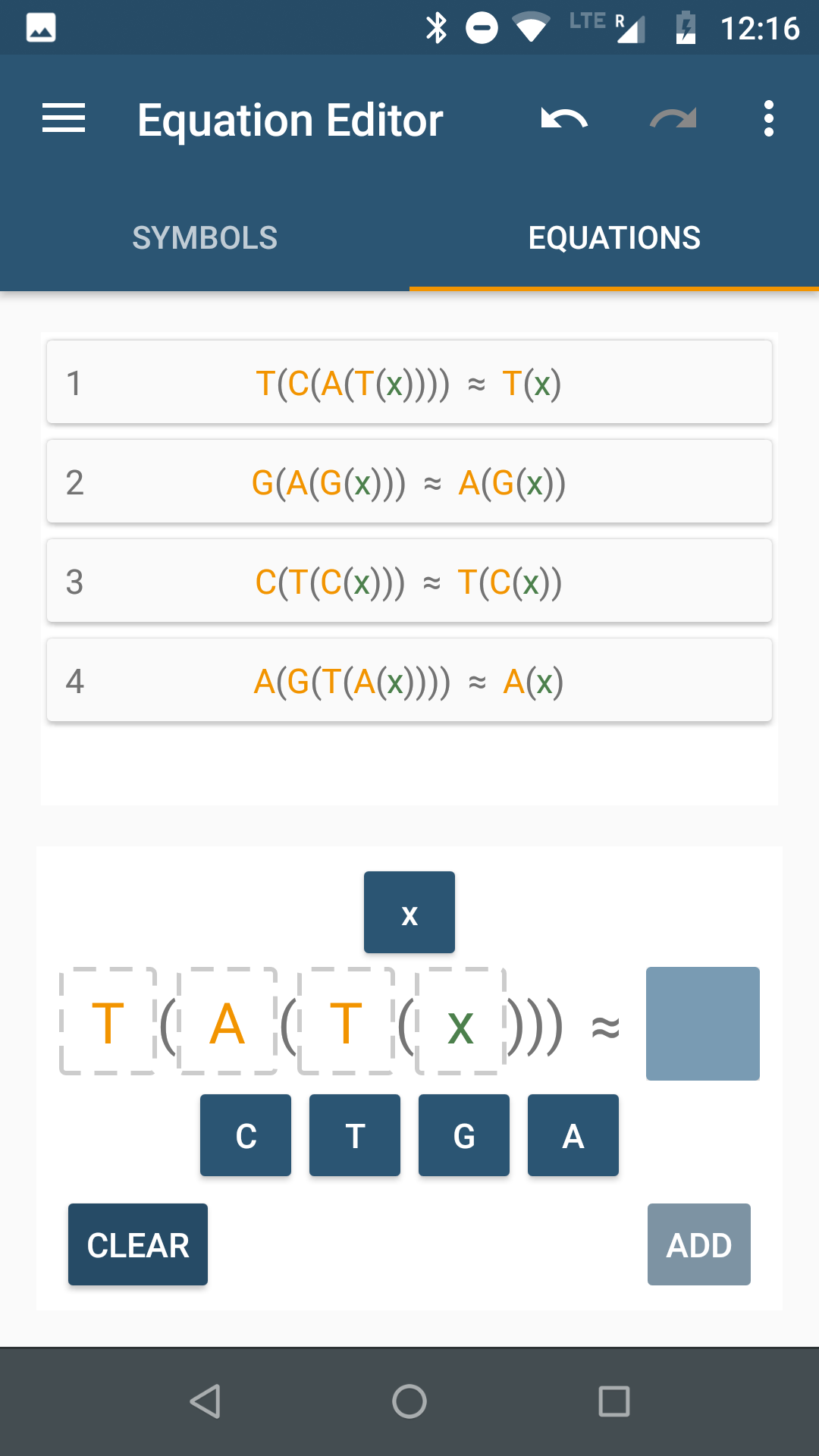}
\end{subfigure}
\qquad
\begin{subfigure}[c]{.6\linewidth}
\includegraphics[width=89mm]{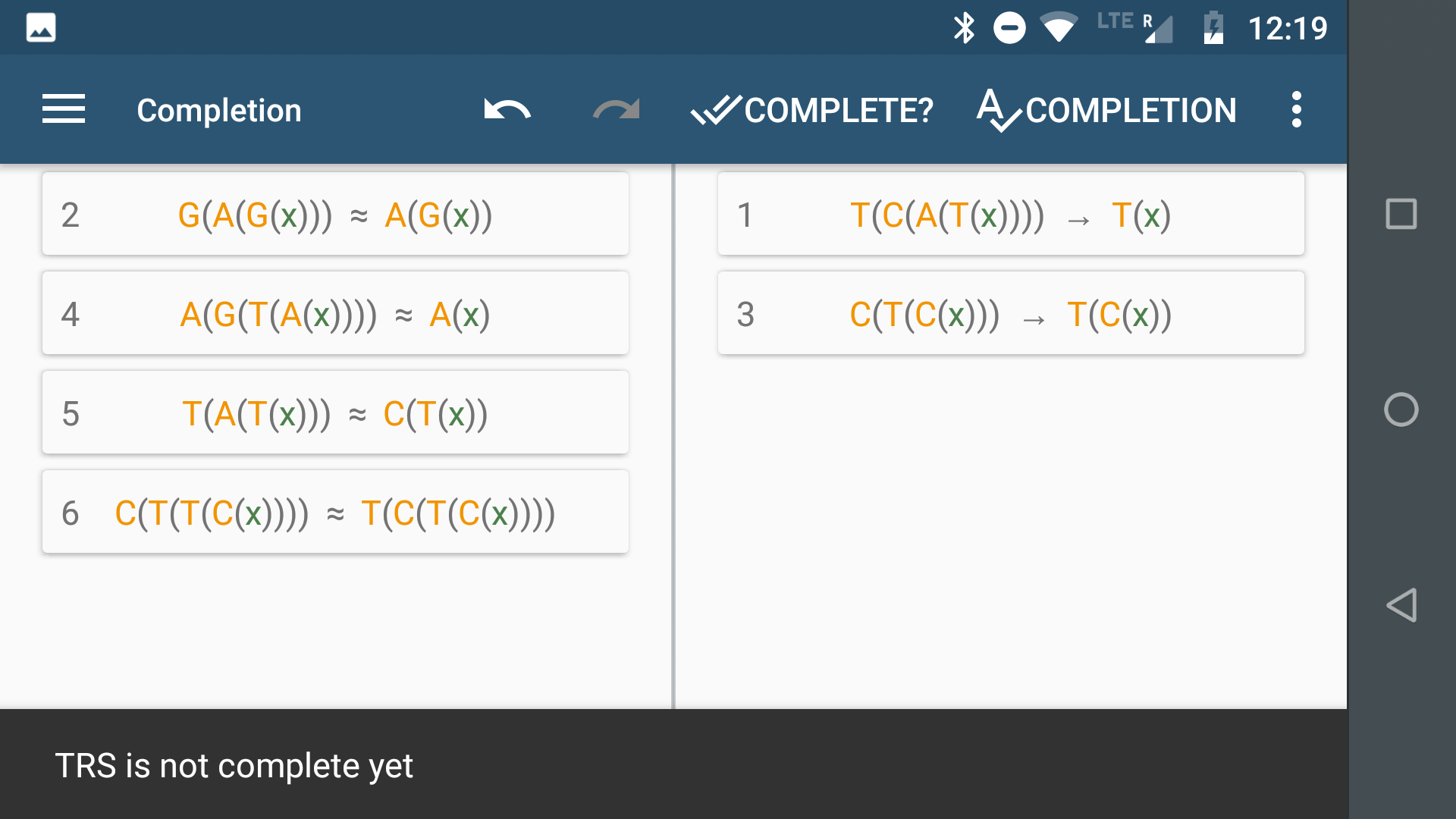}
\end{subfigure}
\caption{The equation editor and the completion interface of the
\KBCV Android application.}
\label{fig:kbcv screenshot}
\end{figure}

Applying completion manually often turns out to be a lengthy
and tedious process, in particular for students, who lack experience.
This observation triggered the development of the
Knuth-Bendix Completion Visualizer (\KBCV)~\cite{SZ12} which is an
implementation of a completion procedure providing two different modes:
In the automatic mode it attempts to complete the system without further
user guidance. But it offers
also an interactive mode, where the user can execute a completion
procedure step-wise, which is useful for students to get acquainted with
completion: All inference rules of the completion inference system
taught in the term rewriting course can be applied separately on the
present equations and rules to observe their effect, and users can also
revert 
steps that turned out to be disadvantageous.
\KBCV is available as a Java executable, via a web interface,
or as an Android application.
\figref{kbcv screenshot} shows screenshots from the \KBCV Android
application run on the gene transformation equations.

\begin{example}
When \KBCV is run in automatic mode on the five equations corresponding to
possible gene transformations in \exaref{genes}, it may produce the TRS
$\RR$ consisting of the following six rules:
\begin{align*}
\m{CT} &\to \m{T} &
\m{TAT} &\to \m{T} &
\m{AGT} &\to \m{AT} &
\m{GA} &\to \m{A} &
\m{ATA} &\to \m{A} &
\m{TCA} &\to \m{TA}
\end{align*}
As $\RR$ is complete, every two convertible terms have a common normal
form. For instance, this is indeed the case for the terms
$\m{TAGCTAGCTAGCT}$ and $\m{CTGACTGACT}$ corresponding to the milk and
cola gene, which confirms that the engineers can perform this
transformation:
\[
\m{TAGCTAGCTAGCT} \,\RabR{!}\, \m{T} \,\LabR{!}\, \m{CTGACTGACT}
\]
The milk gene and the mad cow retrovirus, on the other hand, have
different normal forms:
\[
\m{TAGCTAGCTAGCT} \,\RabR{!}\, \m{T} \neq \m{TGT} \,\LabR{!}\,
\m{CTGCTACTGACT}
\]
Hence there is no danger that an experiment using the above
transformations produces the retrovirus.
\end{example}

The case of the chameleon puzzle in \exaref{chameleons} is more
complicated because the six color-changing rules do not suffice to
model the problem as a TRS as the animals do not meet in a fixed order.
In formal terms, the meeting operator $\cdot$ should be associative and
commutative, i.e., satisfy the following equations:
\begin{align*}
(x \cdot y) \cdot z &\approx x \cdot (y \cdot z) &
x \cdot y &\approx y \cdot x
\end{align*}
However, any completion procedure will fail when confronted with the second
equation since it cannot be oriented into a terminating rewrite
rule. Associative and commutative (AC) operators commonly occur in
practice, for instance in many algebraic specifications.
To deal with such situations, \emph{AC-completion procedures} have been
developed which work \emph{modulo} such equations~\cite{PS81}.
The tool \mkbtt~\cite{WSMK13} offers both a standard and an AC-completion
procedure in an automatic mode, and is available as a binary or via a
web interface.%
\footnote{\url{http://cl-informatik.uibk.ac.at/software/mkbtt/}}

\begin{figure}[tb]
\centering
\includegraphics[width=.98\linewidth]{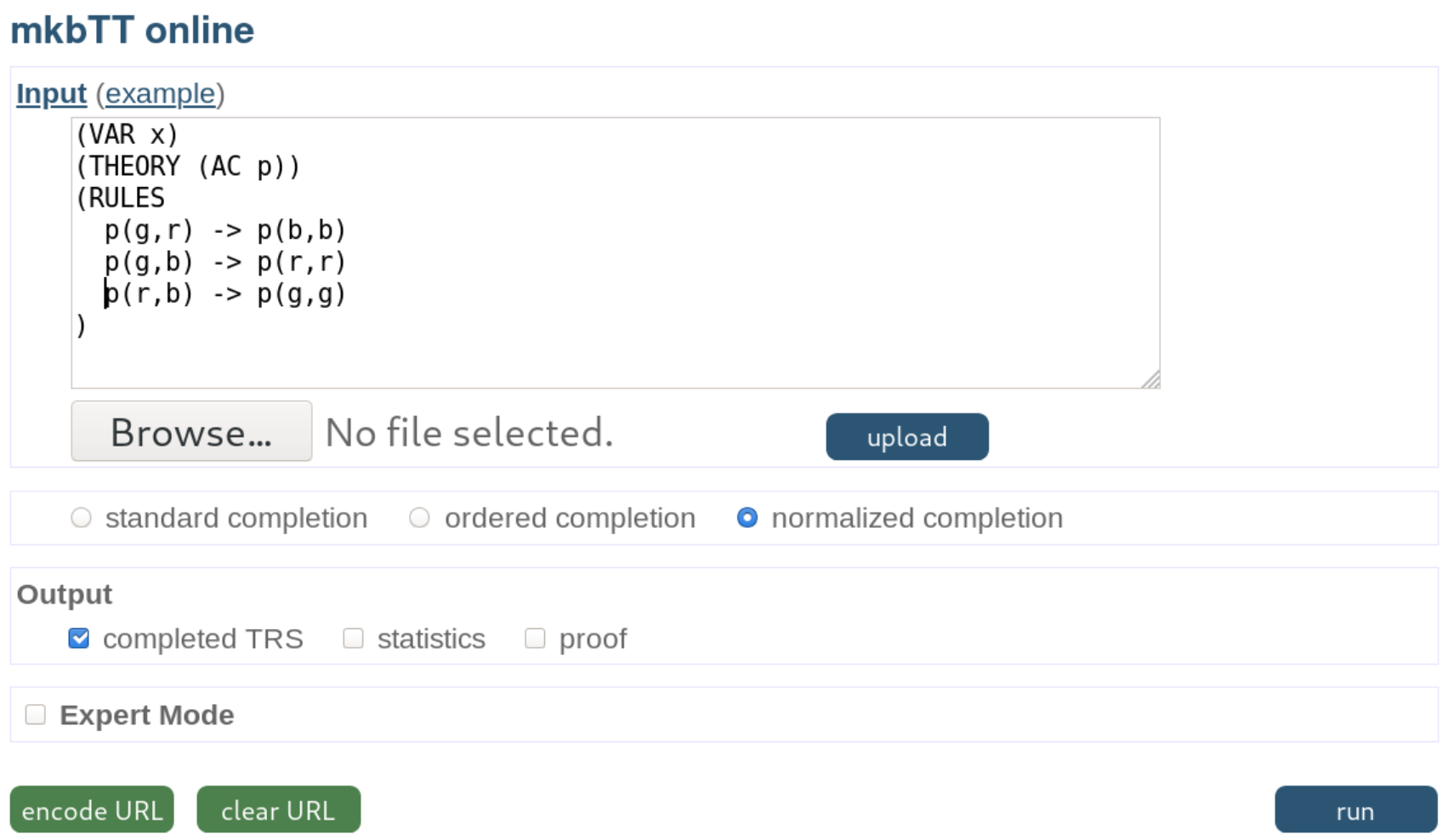}
\caption{The web interface of \mkbtt.}
\label{fig:mkbtt screenshot}
\end{figure}

\begin{example}
When \mkbtt is run on the following three equations with AC operator
$\cdot$:
\begin{align*}
\rC \cdot \gC &\approx \bC \cdot \bC &
\bC \cdot \rC &\approx \gC \cdot \gC &
\gC \cdot \bC &\approx \rC \cdot \rC
\end{align*}
it outputs a TRS $\RR$ that is obtained by reverting one equation and
adding one further rule:
\begin{align*}
\rC \cdot \gC &\to \bC \cdot \bC &
\gC \cdot \gC &\to \rC \cdot \bC &
\gC \cdot \bC &\to \rC \cdot \rC &
\bC \cdot \bC \cdot \bC &\to \rC \cdot \rC \cdot \rC
\end{align*}
This TRS is complete modulo AC. We can now rewrite (modulo AC) the terms
corresponding to the initial colony and 54 monochromatic chameleons to
their respective normal form, where we abbreviate terms of the form
$\rC \cdot \cdots \cdot \rC$ with $n$ occurrences of $\rC$ by $n\:\rC$:
\begin{align*}
20\:\rC \cdot 18\:\bC \cdot 16\:\gC &\,\Rab[\RR/\AC]{!}\,
52\:\rC \cdot 2\:\bC & 54\:\bC &\,\Rab[\RR/\AC]{!}\, 54\:\rC &
54\:\rC &\,\Rab[\RR/\AC]{!}\, 54\:\rC &
54\:\gC &\,\Rab[\RR/\AC]{!}\, 54\:\rC
\end{align*}
Since the normal form  $52\:\rC \cdot 2\:\bC$ of the initial colony is
different from the normal form $54\:\rC$ of 54 monochromatic chameleons
these situations are not convertible. Hence it is impossible that all
animals turn into the same color. By
\href{http://colo6-c703.uibk.ac.at/mkbtt/interface/index.php?problem=%
(VAR\%20x)\%0A(THEORY\%20(AC\%20p))\%0A(RULES\%0Ap(g\%2Cr)\%20-\%3E%
\%20p(b\%2Cb)\%0Ap(g\%2Cb)\%20-\%3E\%20p(r\%2Cr)\%0Ap(r\%2Cb)\%20-\%3E%
\%20p(g\%2Cg)\%0A)\%0A&calculus=normalized}{clicking here}
the interested reader can test the web interface of \mkbtt on this
puzzle. The result is displayed in \figref{mkbtt screenshot}.
\end{example}

\section{First-Order Theory of Rewriting}
\label{sec:first-order theory}

An introductory course on term rewriting typically explains basic
properties like termination and (local) confluence, together with
relationships among these on an abstract level.
For instance, local confluence 
\begin{gather*}
\forall\,s\:\forall\,t\:\forall\,u\:
(s \to t \,\land\, s \to u ~\implies~
\exists\,v\:(t \Ra[*] v \,\land\, u \Ra[*] v))
\end{gather*}
is a strictly weaker property than confluence
\begin{gather}
\forall\,s\:\forall\,t\:\forall\,u\:
(s \Ra[*] t \,\land\, s \Ra[*] u ~\implies~
\exists\,v\:(t \Ra[*] v \,\land\, u \Ra[*] v))
\label{CR}
\end{gather}
and the prototype example of a locally confluent rewrite system that is
not confluent consists of the four rewrite rules
\begin{align*}
\m{a} &\to \m{b} &
\m{b} &\to \m{a} &
\m{a} &\to \m{c} &
\m{b} &\to \m{d}
\end{align*}
involving only constants. This is an example of an abstract rewrite system
(ARS for short), which is a rewrite system over a signature that consists
of constants.

\smallskip

Depending on the application area, one can think of a vast number of 
properties of rewrite systems which are expressible in first-order
formulas like \eqref{CR}. Natural questions arising in this context are
whether a given property $P$ is satisfiable, valid, or implies a different
property $P'$. Such questions also serve as useful exercises in courses
on term rewriting to deepen the understanding of the underlying concepts.
Though for many properties of interest such queries are undecidable,
certain classes of TRSs turn out to admit decision procedures.
Tool support to that end is provided by \tool{FORT}~\cite{RM16,RM18},
an implementation of the decision procedure~\cite{DT90} for the first-order
theory of rewriting for the class of finite left-linear, right-ground
TRSs. This class contains all ARSs. \tool{FORT} has two different modes.

\smallskip

On the one hand, given a left-linear, right-ground TRS and a formula in
the first-order theory of rewriting as input, it decides whether the
property expressed by the formula holds for the given TRS. Formulas are
first-order logic formulas without function symbols and the
predicate symbols include $=$ (equality), $\to$ (one-step rewriting),
$\Ra[*]$ (many-step rewriting), $\Ra[!]$ (rewriting to normal form),
$\PR$ (parallel rewriting), and $\C$ (conversion). Variables in formulas
represent arbitrary ground terms over the signature of the input TRS.
Some of the predicate symbols do not increase the expressive power
of the language but provide convenient shorthands. For instance, $\Ra[!]$
is such a symbol since $s \Ra[!] t$ if and only if
$s \Ra[*] t \,\land\, \lnot\,\exists\,u\:(t \R u)$. For expressing
termination, \tool{FORT} supports the unary predicates $\m{Fin}_R$ for
arbitrary binary regular relations $R$:
\begin{gather*}
\m{Fin}_R(t) \quad\iff\quad
\text{$(t,u) \in R$ for finitely many ground terms $u$}
\end{gather*}
The formula
$\forall\,t\:(\m{Fin}_{\Ra[+]}(t) \,\land\,\lnot\,(t \Ra[+] t))$
states that every term has finitely many reducts and admits no cycle,
which is equivalent to termination for finitely-branching TRSs.

\smallskip

On the other hand, \tool{FORT} provides a synthesis mode in which it
tries to synthesize a left-linear, right-ground TRS that satisfies the
formula given as input. This is practical only when there exists a 
small enough witnessing TRS. For instance, when using \tool{FORT} to
synthesize a locally confluent TRS that is not confluent it delivers
\begin{align*}
\m{g}(\m{g}(x)) &\R \m{g}(\m{g}(\m{g}(\m{c}))) &
\m{g}(\m{g}(\m{g}(\m{c}))) &\R \m{c}
\end{align*}
within a few seconds. We can use the decision mode of \tool{FORT} to
confirm the non-confluence of this TRS. Witness generation, a recent
extension~\cite{RM18}, can be used to find terms in a non-joinable peak:
\begin{align*}
s\colon& \m{g}(\m{g}(\m{g}(\m{g}(\m{c})))) &
t\colon& \m{g}(\m{c}) &
u\colon& \m{c}
\end{align*}
Several input parameters allow to guide the search for a suitable TRS.
We refer to \cite{RM16} for further details.

\smallskip

The current version of \tool{FORT} is written in Java and available as an
executable JAR file.%
\footnote{\url{http://cl-informatik.uibk.ac.at/software/FORT/}}
The decision procedure implemented in \tool{FORT} is based on
tree automata techniques (ground tree transducers, tree automata
operating on encodings of relations on ground terms), which are covered
in a graduate course in Innsbruck on selected topics in term rewriting.
Since tree automata operate on ground terms, the properties that
can be expressed in the first-order theory of rewriting are
properties on ground terms. So the earlier formula \eqref{CR}
stands for ground-confluence, which differs from confluence,
even for left-linear right-ground TRSs. \tool{FORT} provides
special support to deal with non-ground terms for properties
related to confluence. For details we refer to \cite{RM18}.

\section{Strategies}
\label{sec:strategies}

In \exaref{eratosthenes} we have seen an example of a non-terminating
confluent TRS. For terms that have a normal form but also admit infinite
computations, like $\m{take}(\m{s}(\m{s}(\m{0})),\m{primes})$,
it is important to adopt an evaluation strategy that guarantees that
the normal form is reached. The study of strategies has a rich
history---it goes back to the early days of $\lambda$-calculus and
combinatory logic---and many deep results have been obtained
(see \cite{OV03}). Students are typically taught the main strategies and
their normalization behaviour, without going into the proof details.

\newcommand{\ul}[1]{\underline{#1\vphantom{+}}}

\begin{example}
We revisit \exaref{eratosthenes}. If we adopt an eager evaluation strategy
like \emph{leftmost-innermost} in which the leftmost of the innermost
redexes is selected in each reducible term, we will not reach the normal
form of $\m{take}(\m{s}(\m{s}(\m{0})),\m{primes})$,
where we use $\m{n}$ to denote $\m{s}^n(\m{0})$:
\begin{align*}
\m{take}(\m{2},\ul{\m{primes}})
&\,\RbR\, \m{take}(\m{2},\m{sieve}(\ul{\m{from}(\m{2})})) \\
&\,\RbR\, \m{take}(\m{2},\m{sieve}(\m{2}:\ul{\m{from}(\m{3})})) \\
&\,\RbR\, \m{take}(\m{2},\m{sieve}(\m{2}:(\m{3}:\ul{\m{from}(\m{4})}))) \\
&\,\RbR\, \cdots
\intertext{Adopting the \emph{leftmost-outermost} strategy in
which the leftmost of the outermost redexes is selected, the normal form
$\m{2}:(\m{3}:\m{nil})$ is reached:}
\m{take}(\m{2},\ul{\m{primes}})
&\,\RbR\, \m{take}(\m{2},\m{sieve}(\ul{\m{from}(\m{2})})) \\
&\,\RbR\, \m{take}(\m{2},\ul{\m{sieve}(\m{2}:\m{from}(\m{3}))}) \\
&\,\RbR\, \ul{\m{take}(\m{2},\m{2}:
\m{sieve}(\m{filter}(\m{1},\m{from}(\m{3}),\m{1})))} \\
&\,\RbR\, \m{2}:\m{take}(\m{1},
\m{sieve}(\m{filter}(\m{1},\ul{\m{from}(\m{3})},\m{1}))) \\
&\,\RbR\, \m{2}:\m{take}(\m{1},
\m{sieve}(\ul{\m{filter}(\m{1},\m{3}:\m{from}(\m{4}),\m{1})})) \\
&\,\RbR\, \m{2}:\m{take}(\m{1},
\ul{\m{sieve}(\m{3}:\m{filter}(\m{0},\m{from}(\m{4}),\m{1}))}) \\
&\,\RbR\, \m{2}:\ul{\m{take}(\m{1},\m{3}:\m{sieve}(\m{filter}(\m{2},
\m{filter}(\m{0},\m{from}(\m{4}),\m{1}),\m{2})))} \\
&\,\RbR\, \m{2}:(\m{3}:\ul{\m{take}(\m{0},\m{sieve}(\m{filter}(\m{2},
\m{filter}(\m{0},\m{from}(\m{4}),\m{1}),\m{2})))}) \\
&\,\RbR\, \m{2}:(\m{3}:\m{nil})
\end{align*}
\end{example}

Other evaluation strategies like the maximal strategy (previously known
as full-substitution or Gross--Knuth reduction) are more difficult to
apply correctly and this is where the tool \tool{AutoStrat},%
\footnote{\url{http://cl-informatik.uibk.ac.at/software/AutoStrat/}}
comes in handy.
This tool was developed in a bachelor project~\cite{M18} and
also has support for \emph{strategy annotations}. These
were introduced in \cite{vdP01,vdP02} and provide the user with more
control over the evaluation strategy. A key notion here is
\emph{in-time}. We provide an example.

\begin{example}
Consider the TRS consiting of the rewrite rules
\begin{align*}
\alpha\colon ~ x \land \m{T} &\R x &
\gamma\colon ~ \m{T} \lor x &\R \m{T} &
\epsilon\colon ~ \infty &\R \infty &
\beta\colon ~ x \land \m{F} &\R \m{F} &
\delta\colon ~ \m{F} \lor x &\R x
\end{align*}
The Greek letters are used to name to the individual rules. A
strategy annotation specifies for every function symbol the
order in which arguments and potentially matching rewrite rules are
applied. Consider the annotation $A$ with
\begin{align*}
A(\land) &= [2,\alpha,\beta,1] &
A(\lor) &= [1,\gamma,\delta,2] &
A(\infty) &= [\epsilon] &
A(\m{T}) &= A(\m{F}) = [~]
\end{align*}
Suppose we want to evaluate the term
$t = (\infty \land \m{F}) \lor (\m{T} \lor \infty)$.
\begin{itemize}
\item
The strategy annotation $[1,\gamma,\delta,2]$ of its root symbol
$\lor$ tells us that we first look for a redex in the first argument
$\infty \land \m{F}$ of $t$.
\item
The strategy annotation $[2,\alpha,\beta,1]$ for $\land$ indicates to
look for a redex in the second argument $\m{F}$ of
$\infty \land \m{F}$. Since $A(\m{F}) = [~]$, this will fail. So we
discard the first element of $[2,\alpha,\beta,1]$ and try whether rule
$\alpha$ applies. This also fails. Next up is rule $\beta$. Since $\beta$
is applicable, we have found our redex and hence $\infty \land \m{F}$
rewrites to $\m{F}$.
\end{itemize}
So $t$ rewrites to $\m{F} \lor (\m{T} \lor \infty)$.
\end{example}

\smallskip

A strategy annotation defines an evaluation strategy provided the
annotation is \emph{full}, which means that no possibilities are omitted.
The annotation $A$ in the above example is full. If we change
$A(\lor) = [1,\gamma,\delta,2]$ to
$A(\lor) = [\gamma,\delta,2]$ we lose fullness and, as a consequence,
the induced strategy may get stuck on terms which are not yet in
normal form. Indeed, the term $(\m{T} \land \m{T}) \lor \m{F}$ cannot
be reduced since the only redex is in the first argument of $\lor$,
which is excluded from the annotation for $\lor$.

\smallskip

After an annotation-guided rewrite step is performed, the process
starts all over on the resulting term. This typically results in
duplicated efforts to determine the next redex. A function
\emph{normalize} can be defined that continues from the position of the
last step. If the annotation is not only full but also \emph{in-time},
meaning that argument positions are listed before rules that need them,
this function is guaranteed to compute a normal form whenever the
annotation-guided strategy that computes the steps separately is
normalizing. We refer to \cite{vdP01} for formal definitions.

\section{Complexity}
\label{sec:complexity}

While termination is a desirable property, it does not always suffice.
For programs in performance-critical contexts, the computational
complexity is crucial, for simplification processes fast rewriting to
normal form is desired, and frequently the maximal number of rewrite steps
needs to be known for theoretical considerations.
In term rewriting such considerations gave rise to the research area of
complexity. In this section we summarize relevant notions and some
results.

A function symbol $f$ is \emph{defined} in a TRS $\RR$ if $\RR$
contains a rule $f(\seq{\ell}) \to r$. Symbols which are not defined
are \emph{constructor} symbols.
A term $t$ is \emph{basic} with respect to a TRS $\RR$ if
$t = f(\seq{t})$ such that $f$ is defined but none of the arguments
$\seq{t}$ contain any defined symbols.
Complexity analysis in rewriting focuses on the notions
defined below.

\begin{definition}
For a terminating \textup{TRS} $\RR$,
\begin{itemize}
\item
the \emph{derivation height} of a term $t$ is given by $\dh_\RR(t) =
\max\{\,n \mid \text{$t \to_\RR^n u$ for some term $u$}\,\}$,
\item
the \emph{derivational complexity} of $\RR$ is defined as
$\dc_\RR(n) = \max\{\,\dh_\RR(t) \mid |t| = n\,\}$, and
\item
the \emph{runtime complexity} of $\RR$ is defined as $\rc_\RR(n) =
\max\{\,\dh_\RR(t) \mid \text{$t$ is basic and $|t| = n$}\,\}$.
\end{itemize}
Here $|t|$ denotes the size of the term $t$.
\end{definition}

\smallskip

While the derivation height of a term $t$ asks for the maximal number
of rewrite steps that can be performed from $t$ before reaching a normal
form, the derivational complexity of the rewrite system relates the 
derivation height to the size of the starting term.
The runtime complexity restricts this notion to basic terms, which
correspond to potential input of programs. The different concepts are
illustrated by the following example.

\begin{example}
\label{exa:shuffle}
Consider the following TRS $\RR$ representing a functional program to
shuffle a list:
\begin{xalignat*}{3}
\m{nil} \app ys &\to ys &
\m{rev}(\m{nil}) &\to \m{nil} &
\m{shuffle}(\m{nil}) &\to \m{nil} \\
(x \cons xs) \app ys &\to x \cons (xs \app ys) &
\m{rev}(x \cons xs) &\to \m{rev}(xs) \app (x \cons \m{nil})) &
\m{shuffle}(x \cons xs) &\to x \cons \m{shuffle}(\m{rev}(xs)))
\end{xalignat*}
For instance, for the term $t = \m{rev}([1,2])$ we have
$\dh_\RR(t) = 6$ because the following (unique) rewrite sequence to
its normal form has six steps:
\begin{align*}
\m{rev}([1,2])
&\,\RbR\, (\m{rev}([2]) \app [1] \\
&\,\RbR\, (\m{rev}(\m{nil}) \app [2]) \app [1] \\
&\,\RbR\, (\m{nil} \app [2]) \app [1] \\
&\,\RbR\, [2] \app [1] \\
&\,\RbR\, 2 \cons(\m{nil} \app [1]) \\
&\,\RbR\, [2,1]
\end{align*}
Consider a list $xs$ of length $n$. Analysis of the TRS $\RR$ shows that
the number of steps in the (unique) rewrite sequence to normal form is 
\begin{itemize}
\item
linear in $n$ for a term of the form $xs \app ys$,
\item
quadratic for $\m{rev}(xs)$, and 
\item
cubic for $\m{shuffle}(xs)$.
\end{itemize}
A term of the form $\m{shuffle}^n(xs)$ even needs $\OO(n^4)$ steps. Terms
of this shape turn out to witness the worst-case as far as derivation
length in $\RR$ is concerned, hence $\dc_\RR \in \OO(n^4)$.

\smallskip

Note that the term $\m{shuffle}(xs)$ is basic, but
$\m{shuffle}^n(xs)$ is not. Indeed the latter does not correspond to
a run of our shuffle program, where we expect to execute the $\m{shuffle}$
function on some input list.
On the other hand, $\m{shuffle}(xs)$ is basic and a witness
for the cubic runtime complexity of this program, i.e., 
we have $\rc_\RR \in \OO(n^3)$.
\end{example}

\begin{example}
\label{exa:beans3}
We revisit \exaref{beans} from the introduction.
In the case of the TRS $\RR_1$ the number of beans decreases by one with
every move. Hence $\dc_{\RR_1} \in \OO(n)$, i.e., the number of
rewrite steps is linear in the size of the initial configuration.
The TRS $\RR_2$ on the other hand admits very long derivations. The
rewrite sequence
\[
\bB^n(\wB(x)) \to_{\RR_2}
\bB^{n-1}(\wB(\wB(\wB(\bB(x))))) \to_{\RR_2}
\dots \to_{\RR_2}
\wB^{3^n}(\bB^n(x))
\]
shows that $\dc_{\RR_2}$ is exponential.
In fact it is known that a TRS which can be proven terminating by a 
polynomial interpretation has double exponential derivational complexity
in the worst case~\cite{HL89}. If the interpretation is linear as in
\exaref{beans2}, the bound is still single exponential.
\end{example}

\begin{figure}[tb]
\centering
\includegraphics[width=.98\linewidth]{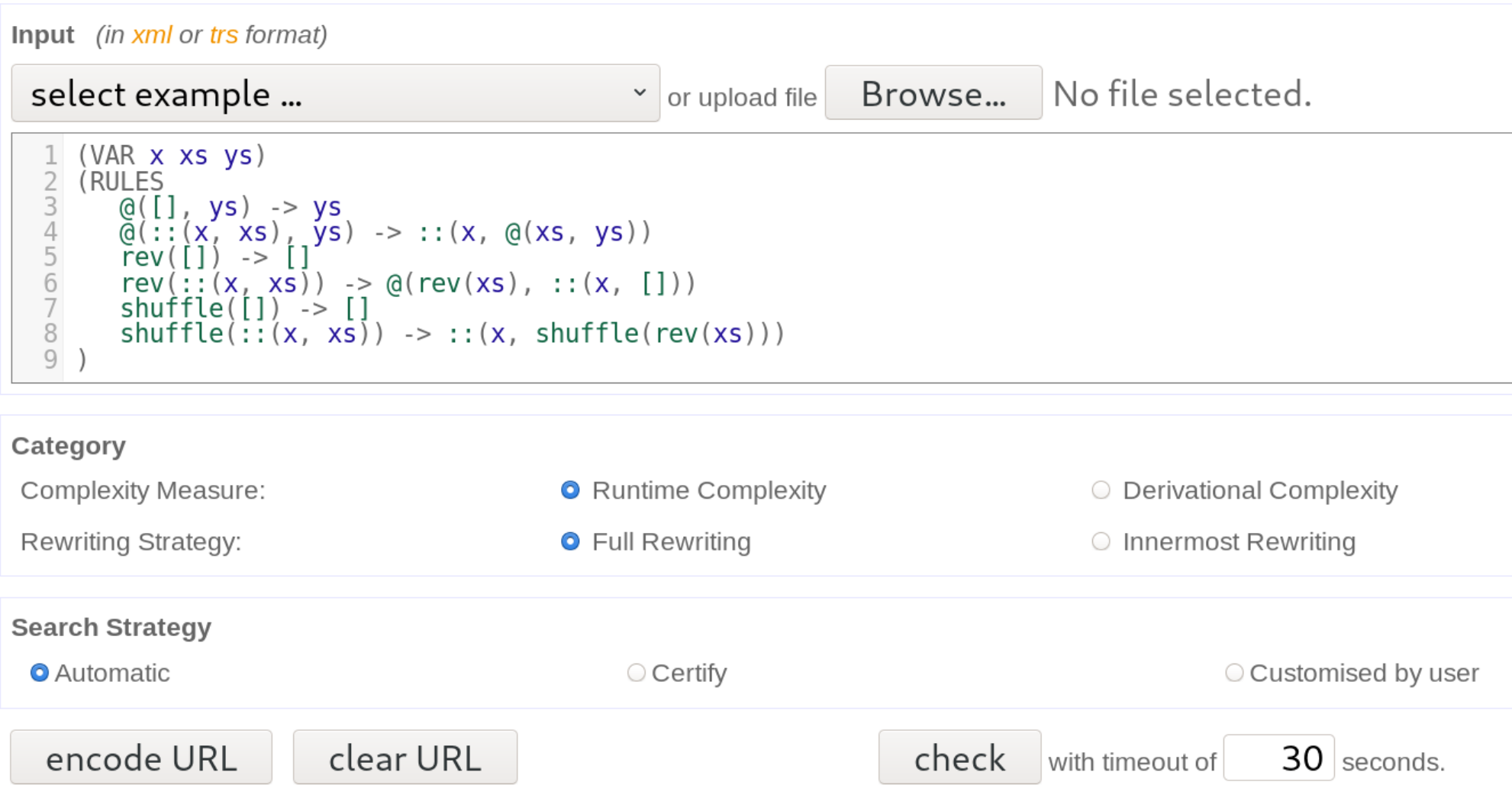}
\caption{The web interface of \TCT.}
\label{fig:tct screenshot}
\end{figure}
The Tyrolean Complexity Tool \TCT is a fully automatic tool for
complexity analysis~\cite{AMS16}.
For instance,
\href{http://colo6-c703.uibk.ac.at/tct/tct-trs/index.php?problem=(VAR\%20%
x\%20xs\%20ys)\%0A(RULES\%20\%0A\%20\%20\%20\%40(\%5B\%5D\%2C\%20ys)\%20-%
\%3E\%20ys\%0A\%20\%20\%20\%40(\%3A\%3A(x\%2C\%20xs)\%2C\%20ys)\%20-\%3E%
\%20\%3A\%3A(x\%2C\%20\%40(xs\%2C\%20ys))\%0A\%20\%20\%20rev(\%5B\%5D)\%%
20-\%3E\%20\%5B\%5D\%0A\%20\%20\%20rev(\%3A\%3A(x\%2C\%20xs))\%20-\%3E\%%
20\%40(rev(xs)\%2C\%20\%3A\%3A(x\%2C\%20\%5B\%5D))\%0A\%20\%20\%20shuffle%
(\%5B\%5D)\%20-\%3E\%20\%5B\%5D\%0A\%20\%20\%20shuffle(\%3A\%3A(x\%2C%
\%20xs))\%20-\%3E\%20\%3A\%3A(x\%2C\%20shuffle(rev(xs)))\%0A)\%09%
\%09&category=rc&strategy=full&search_strategy=webAutomatic_checkbox}%
{clicking here} loads \exaref{shuffle} into
the web interface of \TCT
(with the result shown in \figref{tct screenshot}), and running the
tool on this input establishes
cubic runtime complexity as remarked above (quartic derivational
complexity holds due to the technique from~\cite{Fu19}).
\TCT can not only derive upper bounds on runtime and derivational
complexity of TRSs but also provides resource analysis for Java
bytecode and functional programs, in the latter case also
for higher-order functions, as illustrated by the following example.

\begin{example}
\label{exa:rev}
The following Haskell program reverses a list using the
higher-order function \texttt{fold_left}:
\begin{verbatim}
let rec fold_left f acc = function
   []    -> acc
 | x::xs -> fold_left f (f acc x) xs ;;

let rev l = fold_left (fun xs x -> x :: xs) [] l ;;
\end{verbatim}
\TCT transforms such programs into \emph{higher-order} rewrite systems~%
\cite{TeReSe},
a paradigm whose details are beyond the scope of this paper. Here we 
contend ourselves by noting that \TCT can conclude linear runtime
complexity of this implementation of the \texttt{rev} function, as one
would expect.
\end{example}

\section{Conclusion}
\label{sec:conclusion}

This paper presented automatic tools to analyze term rewrite systems,
developed by the Computational Logic Group at the University of 
Innsbruck. These tools are not only important in research but also
valuable for teaching. In an annual course on term rewriting, as well as
in several editions of the International Summer School on Rewriting,%
\footnote{\url{http://cbr.uibk.ac.at/ifip-wg1.6/summerschool.html}}
they proved highly useful for students as well as teachers to solve and
prepare homework exercises and exam questions.
The tools were mostly developed by (former) graduate students but
also benefitted from student feedback after the use in courses.

\paragraph{Related Work.}
Several other tools support the same TRS analysis tasks as the tools 
described in this paper. In the following paragraphs, we mention
recent tools which are still maintained, and focus on their usability, in
particular via web interfaces since these render them more accessible
to students. We also restrict ourselves to standard TRSs, for special 
types of rewrite systems more tools are available.

In the standard category of the Termination Competition 2019 six tools
participated; ordered by the number of problems solved these are
\tool{AProVE}~\cite{AProVE}, \tool{NaTT}~\cite{NaTT}, 
\TTTT, \tool{mu-term}~\cite{mu-term}, \textsf{Wanda}~\cite{wanda}, and
\textsf{NTI}~\cite{NTI}.
Only \tool{AProVE} and \tool{mu-term} have web interfaces.
The latter allows the user to control (the shape of) polynomial
interpretations, and whether to use RPO and dependency pairs (but neither
LPO, KBO, not matrix interpretations are supported).
In the \tool{AProVE} web interface the user cannot control the strategy 
applied to prove termination (or complexity, which is also supported by
\tool{AProVE}). However, the Java standalone tool offers many options for
control.

In the standard category of the Confluence Competition 2019, besides 
\tool{CSI}, \tool{ACP}~\cite{ACP} and
\tool{CoLL-Saigawa}~\cite{CollSaigawa} participated.
Neither of these has a web interface, but CoCoWeb~\cite{HNM18}
makes them accessible.
Recent completion tools besides \mkbtt and \KBCV are
\tool{maxcomp}~\cite{maxcomp}, and \tool{m\ae{}dmax}~\cite{maedmax}.
Only the latter has a web interface, but it offers few options for control.

To the best of our knowledge, there are no other tools available
which provide similar functionalities as \tool{FORT} or \tool{AutoStrat}.
In proving runtime and derivational complexity of TRSs, the only recent
competitor of \TCT is \tool{AProVE}, already described above.

\paragraph{Outlook.}
Although the presented tools cover by now all approaches explored in 
the basic term rewriting course, the software keeps being extended and
updated to support new techniques emerging from research.
Besides their power as analysis tools, also their
user-friendliness and suitability for teaching can still
be improved. Among possible extensions are web interfaces for \tool{FORT}
and \tool{AutoStrat}, mobile-friendly interfaces for or mobile
applications of other tools besides \KBCV, and more control over the
parameters of proof search, for instance in \tool{CSI} and \TCT.
A single web interface to access all of the tools is another useful
extension. This could include a ``meta-analyzer'' option which uses
the current
tools to analyze
multiple properties of a given TRS at once.

\smallskip

Finally, we comment on the reliability of the presented tools.
Automated reasoning implementations constitute complex pieces of
software due to sophisticated deduction techniques, a high degree of 
optmization, and elaborate heuristics. Hence, implementation errors are
to be expected.
In order to deal with this problem, trusted proof checkers for rewrite
tools have been implemented in the course of the last decade.
To that end, a vast amount of rewriting theory has been formalized
and proved correct
in Isabelle/HOL in the \underline{Isa}belle \underline{Fo}rmalization of 
\underline{R}ewriting (\tool{IsaFoR}) project~\cite{ST14, CETA}.%
\footnote{\url{http://cl-informatik.uibk.ac.at/isafor/}}
From this formalization the proof checker \tool{CeTA} is
generated automatically, which can validate certificates for the 
respective properties (like termination, confluence of completeness)
output by \TTTT, \tool{CSI}, \KBCV, \mkbtt, \tool{FORT}, or \TCT.
Even though many techniques are already supported by
\tool{IsaFoR}/\tool{CeTA}, some of the methods implemented in tools
remain to be added.

\nocite{*}
\bibliographystyle{eptcs}

\providecommand{\noop}[1]{}


\end{document}